\documentclass[aps,prd,floats,floatfix,amssymb,twocolumn,superscriptaddress,nofootinbib]{revtex4-2}

\usepackage{physics}
\usepackage{amsmath}
\usepackage{amsfonts,amssymb}
\usepackage{graphicx,graphics}
\usepackage{hyperref}
\usepackage{ulem}
\usepackage{color}
\usepackage{mathrsfs}
\usepackage{bbold}

\begin{document}

\title{Quantum Approach to Bound States in Field Theory}

\author{Bruno S. Felipe}
\email[]{brunfeli@ifi.unicamp.br}
\affiliation{Instituto de F\'isica Gleb Wataghin, Universidade Estadual de Campinas, 13083-859 Campinas, S\~ao Paulo, Brazil}

\author{Jo\~ao P. M. Pitelli}
\email[]{pitelli@unicamp.br}
\affiliation{Departamento de Matem\'atica Aplicada, Universidade Estadual de Campinas,
13083-859 Campinas, S\~ao Paulo, Brazil}%

\begin{abstract}
It is well known that (possibly non-unique) suitable field dynamics can be prescribed in spacetimes with timelike boundaries by means of appropriate boundary conditions. In Ref.~[J. Math. Phys. {\bf 21}, 2802 (1980)], Wald derived a conserved energy functional for each prescribed dynamics. This conserved energy is related to the positive self-adjoint extensions of the spatial part $A$ of the wave equation $\partial^2\Phi/\partial t^2=-A\Phi$ ($A$ may not be, in principle, essentially self-adjoint). This is quite surprising since the canonical energy is not conserved in these cases. In this paper, we rederive this energy functional from an action principle (with appropriate boundary terms) following Ref.~[Phys. Rev. D, {\bf 69}, 085005, (2004)] and consider field dynamics arising from non-positive self-adjoint extensions of $A$. The spectrum of the resulting theory fails to be positive and unstable mode solutions for classical fields come to light. By studying fields in half-Minkowski spacetime, we illustrate that these unstable classical solutions come as a  consequence of an inverted parabolic potential governing their dynamics. From the quantum mechanical point of view, this leads to an effective inverted harmonic oscillator at the boundary. We then explore these unstable modes behavior, as well as their instabilities, at the quantum level.
\end{abstract} 

\maketitle

\section{\label{sec:intro}Introduction}

In globally hyperbolic spacetimes,  Klein-Gordon field equation (as well as any other linear, second-order hyperbolic system) has a unique solution given initial data $\left.\Phi\right|_\Sigma$ and $n^\mu\nabla_\mu\left.\Phi\right|_\Sigma$ on a Cauchy hypersurface $\Sigma$ normal to the unitary vector $n^\mu$~\cite{wald1}. This is not the case for non-globally hyperbolic spacetimes, where no Cauchy hypersurface can be found.  Nevertheless, it is possible to define at least one suitable field dynamics (given by the so called  Friedrichs extension) on static nonglobally hyperbolic spacetimes with timelike Killing field $\xi=\partial_t$. However, as pointed out by Wald and  Ishibashi~\cite{wald,waldishibashi}, any boundary condition (at the boundary) corresponding to a positive self-adjoint extension of the spatial part of the wave operator $A$ on an appropriate $\mathcal{L}^2$ space, gives rise to a sensible dynamics.

Given any positive self-adjoint extension $A_\gamma$ parametrized by $\gamma$, we can extract a family of complete orthonormal modes solutions $\left\{u_{i}^{\gamma},u_{i}^{\gamma\,\ast}\right\}$ satisfying $\mathsterling_{\xi}u_{i}^{\gamma}=-i\omega u_{i}^{\gamma}$, with $\omega>0$. These modes characterize free states (here, we emphasize the modes dependence on the boundary condition $\gamma$) which spam the field solution as
\begin{equation}\label{phi-positive}
    \phi(t,{\bf x})=\sum_{i}\left[a_{i}u_{i}^{\gamma}(t,{\bf x})+a_{i}^{\dagger}u_{i}^{\gamma\,\ast}(t,{\bf x})\right].
\end{equation}
By imposing the usual equal time commutation relations between $\phi$ and its corresponding conjugated field, we arrive at the usual commutation relations between $a_{i}$ and $a_{i}^{\dagger}$. These turn out to be operators acting on an appropriate Fock space with the vacuum state  $\ket{0}$ satisfying
\begin{equation}\label{vacuum}
    a_{i}\ket{0}=0,\quad \forall i.
\end{equation}

In this paper, we study the quantization of the Klein-Gordon field on half-Minkowksi spacetime ($z>0$) satisfying (at $z=0$) a boundary condition corresponding to a non-positive self-adjoint extension of $A$. As a result, modes with imaginary energy, i.e., $\textrm{Im}(\omega)\neq 0$ give rise to unstable dynamics. Furthermore, the decomposition into positive and negative frequencies given by Eq.~(\ref{phi-positive}) is meaningless for this class of mode solutions. Hence, the usual quantization procedure based on the construction of a Fock space with $\ket{0}$ as its vacuum state breaks down.

\begin{figure}[htb!]
    \centering
    \includegraphics[scale=1]{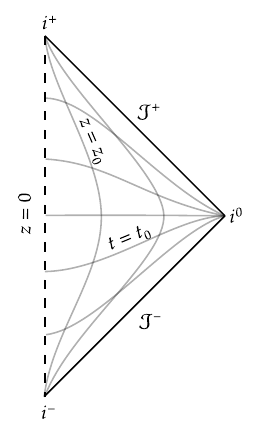}
    \caption{Conformal diagram of the 2-dimensional half-Minkowski space with timelike boundary at $z=0$.}
    \label{conformal:mink}
\end{figure}

The aim of this paper is to study and interpret this pathological bound state solution. We will restrict our analysis to the half-Minkowski spacetime $\overset{\circ}{\mathbb{H}}$\footnote{We point out that $\overset{\circ}{\mathbb{H}}$ is conformal to anti-de Sitter spacetime in Poincaré coordinates ($\mathrm{PAdS}$). Hence, our results fit equally well to a conformal Klein-Gordon field in $\mathrm{PAdS}$.}. This spacetime is described by the line element
\begin{equation}\label{half-mink}
    ds^2=g_{\mu\nu}dx^{\mu}dx^{\nu}=-dt^2+dx^2+dy^2+dz^2,
\end{equation}
where $t,x,y\in \mathbb{R}$ and $z\in \mathbb{R}^{+}$. Suppressing the coordinates $x$ and $y$, its conformal diagram is given in Fig.~\ref{conformal:mink}, where we see that the ``wall'' $z=0$ affects every event on the spacetime bulk.
The appropriate boundary conditions at $z=0$, i.e., those for which $A_\gamma$ is self-adjoint are the usual Robin boundary conditions (RBC). In Ref.~\cite{saharian}, Saharian showed that these boundary conditions can be extracted from a variational principle with appropriate boundary terms in the action $S=S_{\textrm{bulk}}+S_{\textrm{boundary}}$. This generalized action leads to a conserved energy $E=E^{\textrm{bulk}}+E^{\textrm{surface}}$ which turns out to be equivalent to Wald's energy~\cite{wald}. 

When the extension $A_\gamma$ fails to be positive, the wave equation also has a ``bound state'' $\phi_b(t,{\bf x})=\chi(x,y)\psi_\gamma(z)e^{(\pm i \textrm{Re}(\omega)\mp \textrm{Im}(\omega))t}$, with 
\begin{equation}
\int_0^{\infty}{|\psi_\gamma(z)|^2\mathrm{d}z}<\infty.
\end{equation}
In Ref.~\cite{martinez}, a zero-mode solution $\phi(t,\theta)=q(t)$ derived from a non-usual (Neumann) boundary condition on a cylindrical spacetime was studied. By incorporating this solution back into the total action, Martin-Martinez et al showed that the resulting Lagrangian for $q(t)$ was equivalent to a free particle one. This zero-mode component was (first) quantized in a corresponding one-particle non-relativistic Hilbert space. 

Following the procedure of the zero-mode solution given in Ref.~\cite{martinez}, we also incorporate the bound state solution back into the generalized action. As a result, we find that the surface action introduces a (inverted) parabolic potential for the time domain $\chi(t)$ of the bound state. We interpret the resulting Lagrangian for $\chi(t)$ as an inverted harmonic oscillator. By extending the concept of operators in quantum mechanics from Hilbert space to the so-called Rigged Hilbert space, we establish a ground state for our bound state and interpret it at the quantum level.

This paper is organized as follows. In section \ref{sec:action}, we introduce the action formalism with a surface term to address the field dynamics in half-Minkowski spacetime. We demonstrate how the RBC give rise to unstable modes (bound states), governed by an {\it inverted} harmonic oscillator-like potential. In section \ref{sec:InvertedOscillator}, we review the main characteristics of the quantum inverted harmonic oscillator, and then proceed to quantize the bound field using a direct analogy with the canonical quantization procedure of free fields. In the following section, \ref{sec:stresstensor}, we show that this approach is consistent with energy conservation in both classical and quantum scenarios. Finally, our concluding remarks are presented in section \ref{sec:conclusion}.

\section{\label{sec:action}Field solution from the action with a boundary term}
Let $\Phi:\overset{\circ}{\mathbb{H}} \to \mathbb{R}$ be a real massless scalar field in half-Minkowski space defined by the line element~(\ref{half-mink}). At the ``wall'' $z=0$, one can define a smooth surface $\partial\overset{\circ}{\mathbb{H}}$ with the induced metric $h_{\mu\nu}=\text{diag}(-1,1,1)$ and normal to the unitary vector $n^{\mu}=\delta_{z}^{\mu}$. Then, following Ref.~\cite{saharian}, we can construct the total action with both bulk and surface terms in the form
\begin{equation}\label{total-action}
    S[\Phi]=-\frac{1}{2}\int_{\overset{\circ}{\mathbb{H}}}\mathrm{d}^4x\,\partial^{\mu}\Phi\partial_{\mu}\Phi+\frac{\gamma}{2}\int_{\partial\overset{\circ}{\mathbb{H}}} \mathrm{d}^3x\,\Phi^2.
\end{equation}
Here,  $\gamma\neq0$ represents a mass parameter of the field at the surface $\partial\overset{\circ}{\mathbb{H}}$.

Taking the variation of $S$ with respect to the field, we obtain
\begin{equation}
    \begin{aligned}
        \delta_{\Phi}S&=\int_{\overset{\circ}{\mathbb{H}}}\mathrm{d}^4x\left(\partial^{\mu}\partial_{\mu}\Phi\right)\delta\Phi+\gamma\int_{\partial\overset{\circ}{\mathbb{H}}}\mathrm{d}^3x \left(\Phi+\frac{1}{\gamma}\Phi^{\prime}\right)\delta\Phi,
    \end{aligned}
\end{equation}
where we have integrated by parts and used Stoke's theorem. For any $\delta \Phi$ we have $\delta_{\Phi}S=0$ iff
\begin{equation}\label{EOM}
    \partial_{\mu}\partial^{\mu}\Phi=0\quad \text{and}\quad \left( \Phi+\frac{1}{\gamma}\Phi^{\prime}\right )\bigg|_{z=0} =0,
\end{equation}
with $\Phi^{\prime}$ denoting the field derivative with respect $z$, i.e., $\Phi^{\prime}\equiv\partial_z\Phi$. Notably, both the equation of motion and the Robin boundary condition are derived through the variation principle from the action with the surface term (\ref{total-action}). In this notation, the Robin boundary condition recovers the Dirichlet boundary condition ($\left.\Phi\right|_{z=0}=0$) and the Neumann boundary condition ($\left.\Phi^{\prime}\right|_{z=0}=0$) when $\gamma\to \infty$ and $\gamma \to 0$, respectively.

Writing $u(t,{\bf x})=\chi(t)X(x)Y(y)\psi(z)=e^{-i \omega t}e^{ik_x x}e^{ik_y y}\psi(z)$, the wave equation (\ref{EOM}) yields
\begin{equation}\label{sep-variable}
    \begin{aligned}
        -\frac{1}{\psi}\frac{\mathrm{d}^2\psi}{\mathrm{d}z^2}&=-\frac{1}{\chi}\frac{\mathrm{d}^2\chi}{\mathrm{d}t^2}+\frac{1}{X}\frac{\mathrm{d}^2X}{\mathrm{d}x^2}+\frac{1}{Y}\frac{\mathrm{d}^2Y}{\mathrm{d}y^2}\\
        &=\omega^2-k_x^2-k_y^2\\
        &=\omega^2-{\bf k}^2=q^2,
    \end{aligned}
\end{equation}
which for the $z$-coordinate can be understood as a standard Sturm-Liouville problem~\cite{dappiaggi} where the eigenvalue is denoted as $\lambda\equiv q^2$, subject to the Robin boundary condition at $z=0$. In the case of a positive eigenvalue, $\lambda=q^2>0$, the $z$-component solution is expressed as a linear combination of the linearly independent base solutions $\left\{\sin(qz),\cos(qz)\right\}$, given by
\begin{equation}
    \psi_q(z)=A \sin(qz)+B\cos(qz).
\end{equation}
This solution satisfies the boundary condition (\ref{EOM}) if $B=-Aq/\gamma$.  The normalized mode solution takes the form
\begin{equation}
    u^{\gamma}_{{\Vec{k}}}(t,{\bf x})=\frac{\gamma e^{i{\bf k}\underline{{\bf x}}-i\omega t}\left(\sin(qz)-\frac{q}{\gamma}\cos(qz)\right)}{2\pi^{\frac{3}{2}}\sqrt{\omega \gamma^{2}+\omega q^{2}}},
\end{equation}
with $\Vec{k}=(q,{\bf k})=(q,k_x,k_y)$, $\omega=\sqrt{q^2+ {\bf k}^2}$ and $\underline{\bf x}=(x,y)$.

For negative values of $q^2$, an alternative linearly independent solution satisfying the Robin boundary condition emerges when $\lambda=-q^{2}=\gamma^2$. This implies $q=\pm i \gamma$, resulting in the $z$-component adopting a real exponential form. This real exponential can be normalized in $\mathcal{L}^2(\mathbb{R}_{+},dz)$ to yield $\psi_{\gamma}(z)=\sqrt{2\gamma}\exp(-\gamma z)$. The normalized solution for modes with purely imaginary values of $q$ can then be expressed as
\begin{equation}\label{imaginary-solution}
    u^{(\text{im})}_{\gamma}(t,{\bf x})=\sqrt{\frac{\gamma}{\omega}}\frac{e^{i {\bf k}  \underline{{\bf x}} -i \omega t-\gamma z}}{2\pi},
\end{equation}
where $\omega=\sqrt{{\bf k}^2-\gamma^2}$. However, it is important to note that for these states, two distinct kinds of solutions exist. For values of ${\bf k}$ such that $|{\bf k}|>\gamma$, the modes
\begin{equation}
    u_{i}^{\gamma}(x)=\left(u_{\Vec{k}}^{\gamma},u^{(\text{im})}_{\gamma<|{\bf k}|}\right)
\end{equation}
are  eigenvectors of the Killing field $\partial_t$ with corresponding eigenvalues $-i\omega$ where $\omega>0$. Therefore, the set $\left\{u_{i}^{\gamma},u_{i}^{\gamma\ast}\right\}$ can form a basis for the ``free'' field in the structure of Eq.~(\ref{phi-positive}), allowing for the standard quantization procedure.

When $|{\bf k}|<\gamma$, the frequency becomes purely imaginary, $\omega=i\sqrt{\gamma^2-{\bf k}^2}$, leading the solution (\ref{imaginary-solution}) to be time-divergent for larger values of $t$. Additionally, the decomposition (\ref{phi-positive}) loses its interpretation, and no vacuum states can be associated with these modes at the quantum level. To overcome this situation, we will avoid solving explicitly its time dependence $\chi(t)$ and express the total classical solution as
\begin{equation}
    \Phi(t,{\bf x})=\phi(t, {\bf x})+\phi_{b}(t, {\bf x}),
\end{equation}
where $\phi(t,{\bf x})$ represents the {\it free field} written in the form of Eq.~(\ref{phi-positive}) for the modes $u_i^{\gamma}$, while 
\begin{equation}\label{phi_b1}
    \phi_b(t, {\bf x})=\sqrt{2\gamma}e^{-\gamma z}\int_{|{\bf k}|<\gamma} \frac{\mathrm{d}^2{\bf k}}{2\pi}e^{i {\bf k} \underline{{\bf x}}}\chi_{{\bf k}}(t),
\end{equation}
defines what we call the {\it bound state field} solution, characterized by the integration over all bound states $u_{\gamma>|{\bf k}|}^{(\text{in})}$.

In order to find the precise potential that causes the time-divergence of the bound field, let's reintroduce $\phi_b(t,{\bf x})$ into the total action (\ref{total-action}) and perform the spatial integration as follows (we denote the time derivative by a dot)
\begin{widetext}
    \begin{equation}\label{action-oscilator}
    \begin{aligned}
        S[\phi_b]&=-\frac{1}{2}\int_{t_1}^{t_2} \mathrm{d}t\int \mathrm{d}^{3} x \left[-\dot{\phi}_b^2+\phi_b^{\prime\,2}+(\partial_x\phi_b)^2+(\partial_y\phi_b)^2\right]+\frac{\gamma}{2} \int_{t_1}^{t_2}\mathrm{d}t\int\mathrm{d}^2x \phi_b^2\\
        &=\int_{t_1}^{t_2} \mathrm{d}t\int \mathrm{d}^2{\bf k} \left[\frac{\dot{\chi}_{{\bf k}}\dot{\chi}_{{\bf k}}^{\ast}}{2}+\left(\gamma^2-{\bf k}^2\right)\frac{\chi_{{\bf k}}\chi_{{\bf k}}^{\ast}}{2}\right],\quad {|{\bf k}|<\gamma}.
    \end{aligned}
\end{equation}
\end{widetext}

By recovering the conventional Lagrangian expression as $L=T-V$ and defining $\omega_{\bf k}^2\equiv \gamma^2-{\bf k}^2$, the last line of the above equation shows that the behavior of $\chi_{{\bf k}}$ for the bound states resembles a massive particle subjected to the potential $V=-\omega_{\bf k}^2|\chi_{{\bf k}}|^2/2$ - often referred to as the {\it inverted harmonic oscillator} (IHO). Essentially, the bound field $\phi_b(t,{\bf x})$ evolves in time as a 
 collection of inverted harmonic oscillators satisfying $|{\bf k}|<\gamma$. Moreover, in Eq.~(\ref{action-oscilator}), we can find the physical origin of the time-divergence issue. The bulk term generates the potential of a standard harmonic oscillator $(\gamma^2+{\bf k}^2)|\chi_{\bf k}|^2/2$, while the surface action generates the term $-\gamma^2|\chi_{\bf k}|^2$. For values of $|{\bf k}|>\gamma$, the combined potentials result in standard oscillators with dislocated frequencies (characterizing the modes $u_{\gamma<|{\bf k}|}^{(\text{im})}$). Conversely, for values $|{\bf k}|<\gamma$, the potential coming from the surface exceeds the bulk contribution, giving rise to the inverted harmonic oscillator behavior.

It is worth noting that the case $V=0$, which implies $\omega_{\bf k}^2=\gamma^2={\bf k}^2=0$, corresponds to the Neumann boundary condition. This results in a non-relativistic free particle behavior, namely the zero-mode solution (as extensively discussed in Ref.~\cite{martinez}). For the general case of $V\neq 0$, the system consistently exhibits an IHO behavior — even in the simplest scenario of a bi-dimensional spacetime where ${\bf k}=0$ and $\omega=\gamma$. In this paper, we focus on investigating states where the time domain manifests this unconventional dynamics of an inverted harmonic oscillator. As the potential $V$ is unbounded from below, classical solutions suggest time divergence when particles interact with this parabolic barrier. However, as elaborated in the subsequent section, at the quantum level, subtle intricacies emerge, providing an opportunity for a profound understanding of this peculiar phenomenon.

\section{The bound state field as Inverted Harmonic Oscillator}\label{sec:InvertedOscillator}

Let us first summarize the quantum aspects of the inverted harmonic oscillator, as this system is not commonly covered in standard textbooks. Subsequently, we will apply the same quantization techniques used for the IHO to the bound state field in a manner consistent with the canonical quantization procedure. 

The starting point is the Hamiltonian
\begin{equation}\label{hamiltonian}
    H_{\text{IHO}}=\frac{p^2}{2}-\frac{\widetilde{\omega}^2 x^2}{2},
\end{equation}
which corresponds to the Hamiltonian of a standard harmonic oscillator with its frequency changed as $\omega\to \pm i \widetilde{\omega}$. Here, $\widetilde{\omega}$ is understood as the decay rate, while $x$ is a generalized coordinate, and $p$ is its conjugate momentum. In the theory of IHO, various basis, and consequently,  various interpretations for the above Hamiltonian, can be used (e.g, $H_{\text{IHO}}$ can be understood as a {\it generator of squeeze} in quantum optics~\cite{scully}, as well describe {\it resonant states}~\cite{oscilador1,oscilador2} or even a way to quantize {\it damped systems}~\cite{dariusz1,dariusz2}). Each choice of basis allows us to extract specific meanings from the system, and in the context of our analysis in field theory, we emphasize the importance of two distinct bases as follows.

\subsection{Energy eigenstates of the IHO}

The quantum formulation of the IHO can be accomplished by directly imposing the canonical commutation relation $\left[x,p\right]=i$ (using $\hslash =1$), ensuring that the Hamiltonian (\ref{hamiltonian}) represents a self-adjoint operator in the Hilbert space $\mathcal{H}$. Its energy eigenstates, denoted as $\ket{X^{\mathscr{E}}_{\pm}}$, are doubly degenerated in the eigenvalue $\mathscr{E}\in\mathbb{R}$, satisfying
\begin{equation}
H_{\text{IHO}} \ket{X^{\mathscr{E}}_{\pm}}=\mathscr{E}\ket{X^{\mathscr{E}}_{\pm}},
\end{equation}
where the symbols $\pm$ indicate the analogy with in(out)-scattering states.

Realizing the Hilbert space in $x$, i.e., $\mathcal{H}=\mathcal{L}^2(\mathbb{R}_{x})$, the above eigenvalue equation becomes 
\begin{equation}\label{hamiltonian2}
\frac{\mathrm{d}^2X_{\pm}^{\mathscr{E}}}{\mathrm{d}x^2}+\left(\widetilde{\omega}^2x^2+2\mathscr{E}\right)X_{\pm}^{\mathscr{E}}=0,
\end{equation}
with $X_{\pm}^{\mathscr{E}}(x)=\bra{x}\ket{X^{\mathscr{E}}_{\pm}}$. The solution to Eq.~(\ref{hamiltonian2}) is provided by a linear combination of parabolic cylinder functions along with an appropriate boundary condition. In our case, and extremely important for future definition, a suitable solution (discussed in Ref.~\cite{dariusz2}) can be written as
\begin{equation}
    X_{\pm}^{\mathscr{E}}(x)=\frac{C_0}{\sqrt{2\pi\widetilde{\omega}}}\sqrt{i}^{\nu+1/2}\Gamma(\nu+1)D_{-\nu-1}(\mp\sqrt{-2i\widetilde{\omega}}\,x),
\end{equation}
where we identify
\begin{equation*}
    C_0=\left(\frac{\widetilde{\omega}}{2\pi^2}\right)^{1/4}\quad\text{and}\quad \nu=-\left(i\frac{\mathscr{E}}{\widetilde{\omega}}+\frac{1}{2}\right).
\end{equation*}

Furthermore, there is another linearly independent base solution for Eq.~(\ref{hamiltonian2}), identified by the conjugate elements of $X_{\pm}^{\mathscr{E}}$. By considering the conjugation relationships $$\overline{\nu+1}=-\nu \quad \text{and}\quad \overline{\sqrt{i}^{\nu+1/2}}=\sqrt{i}^{\nu+1/2},$$ we can express the conjugate states $Y_{\pm}^{\mathscr{E}}(x)=\overline{X_{\pm}^{\mathscr{E}}(x)}$ as 
\begin{equation}\label{X-conjugated}
    Y_{\pm}^{\mathscr{E}}(x)=\frac{C_0}{\sqrt{2\pi\widetilde{\omega}}}\sqrt{i}^{\nu+1/2}\Gamma(-\nu)D_{\nu}(\mp\sqrt{2i\widetilde{\omega}}\,x).
\end{equation}
In other words, these new states define the energy eigenstate  $H_{\text{IHO}}\ket{Y_{\pm}^{\mathscr{E}}}=-\mathscr{E}\ket{Y_{\pm}^{\mathscr{E}}}$ $-$ a direct consequence of $\nu+1\to-\nu$ which corresponds $\mathscr{E}\to-\mathscr{E}$.

The fundamental point to observe here is that the states $X^{\mathscr{E}}_{+}$ and $X^{\mathscr{E}}_{-}$ satisfy 
\begin{equation}
    \int_{-\infty}^{\infty}\overline{X^{\mathscr{E}}_{\pm}(x)}X^{\mathscr{E}^{\prime}}_{\pm}(x)\mathrm{d}x=\delta(\mathscr{E}-\mathscr{E}^{\prime}),
\end{equation}
which identify their non-square normalized nature. Since this family of four states $X^{\mathscr{E}}_{+},\, X^{\mathscr{E}}_{-},\, Y^{\mathscr{E}}_{+}$ and $Y^{\mathscr{E}}_{-}$ are not normalizable, they cannot be elements of the Hilbert space $\mathcal{H}$. Instead, they are distribution functions belonging to the dual Schwartz space $\mathcal{S}(\mathbb{R}_{x})^{\times}$, which can be defined through the Gelfand triplet (also known as rigged Hilbert space)
\begin{equation}\label{triplet1}
    \mathcal{S}(\mathbb{R}_{x})\subset \mathcal{L}^{2}(\mathbb{R}_{x})\subset \mathcal{S}(\mathbb{R}_{x})^{\times},
\end{equation}
where $\mathcal{S}$ denotes the Schwartz space - the space of test functions. In other words, the quantum prescription of the bound field requires an enlargement of the Hilbert space $\mathcal{H}$ to the space $\mathcal{S}^{\times}$ to accommodate the IHO energy eigenstates. The triplet of spaces defined above can be precisely read in the following way: given a Hilbert space $\mathcal{H}$ with topology $\tau_{\mathcal{H}}$, we identify a subspace $\mathcal{S}$ with a stronger topology $\tau_{\mathcal{S}}$ compared to $\tau_{\mathcal{H}}$. Subsequently, we define its dual $\mathcal{S}^{\times}$ which is endowed with $\tau_{\mathcal{S}^{\times}}$, a topology weaker than $\tau_{\mathcal{H}}$, in order to characterize the space of energy, now called {\it generalized eigenvectors}, i.e., $\ket{X^{\mathscr{E}}_{\pm}} \in \mathcal{S}^{\times}$. Additionally, as extensively discussed by A. Bohm~\cite{bohm1,bohm2,bohm3,bohm4}, the Dirac formalism of {\it brackets} works completely the same way within the Gelfand triplet\footnote{It is essential to recognize that different types of endowed topology yield different triplets. Therefore, identifying $\mathcal{S}$ as the Schwartz space is specific to the case of IHO. For various examples and a comprehensive introduction to rigged Hilbert spaces, we refer to~\cite{Madrid1,Madrid2}.}, with the interpretation that the {\it bra} elements belong to the smallest space, $\mathcal{S}$, such that it forces the {\it braket} operation to converge.

\subsection{Decay and growth states}

Similar to the standard harmonic oscillator, where the usual coordinate transformation $(x, p) \to (a, a^{\dagger})$ is performed, a comparable procedure can be applied in the context of the inverted harmonic oscillator. One can perform a canonical transformation of $x$ and $p$ to define the new operators as $b^{\pm}:=\sqrt{\frac{\widetilde{\omega}}{2}}\left(x\pm \frac{1}{\widetilde{\omega}} p\right)$. Now, the commutation relation for $b^{+}$ and $b^{-}$ writes
\begin{equation}\label{commutation}
    \left[b^{+},b^{-}\right]=-i \quad\text{and}\quad \left[b^{\pm},b^{\pm}\right]=0.
\end{equation}
Here, $b^{\pm}$ are both essentially self-adjoint operators in the Schwartz space $\mathcal{S}$~\cite{shimbori2}. Expressing the Hamiltonian (\ref{hamiltonian}) in terms of these new coordinates, we get
\begin{equation}\label{hamiltonian3}
    H_{\text{IHO}}=-\frac{\widetilde{\omega}}{2}\left(b^{+}b^{-}+b^{-}b^{+}\right).
\end{equation}

Then, we can define the {\it ground states} $\ket{f_0^{\pm}}$, such that
\begin{equation}\label{groundstate}
    b^{\mp}\ket{f_0^{\pm}}=0,
\end{equation}
meaning that $b^{+}$ annihilates $\ket{f_0^{-}}$ and $b^{-}$ annihilates $\ket{f_0^{+}}$. With these states in hands, it becomes possible to construct new states by applying $b^{\pm}$ $n$-times to their respective non-annihilated states, defining $\ket{f_{n}^{\pm}}=(b^{\pm})^{n}\ket{f_{0}^{\pm}}$. The action of the Hamiltonian (\ref{hamiltonian3}) on these states (by direct using the commutation relations (\ref{commutation})) results in
\begin{equation}\label{eigenstate}
    H_{\text{IHO}} \ket{f_n^{\pm}}=\mp E_{n}\ket{f_n^{\pm}},
\end{equation}
where 
\begin{equation}
    E_n=i\,\widetilde{\omega}\left(n+1/2\right),\quad n \in \mathbb{R}^{+}_{0}.
\end{equation}
This implies that $\ket{f_n^{\pm}}$ are generalized eigenvectors of $H_{\text{IHO}}$ with complex eigenvalues $E_n$ — a direct consequence of the potential in $H_{\text{IHO}}$ not being bounded from below.

The nature of these states can be understood in terms of representatives. For $f_n^{\pm}(x)=\bra{x}\ket{f_n^{\pm}}$ we find (see Refs.~\cite{dariusz2} and~\cite{shimbori2} for details)
\begin{equation}
    f_{n}^{\pm}(x)=N_{n}^{\pm}e^{\mp i \widetilde{\omega}x^2/2}H_{n}(\sqrt{\pm i \widetilde{\omega}}\,x),
\end{equation}
where $N_{n}^{\pm}$ is a normalization constant and $H_{n}$ stands for the $n$th Hermite polynomial. Therefore, these solutions are tempered distributions, i.e., they are not elements of the Hilbert space $\mathcal{L}^2(\mathbb{R}_x)$ but they do belong to the dual of the Schwartz space $\mathcal{S}(\mathbb{R}_x)^{\times}$. To distinguish the spaces of $f_n^{+}$ and $f_n^{-}$, one usually introduces two dual of the Schwartz spaces denoted as $\mathcal{S}_{\pm}(\mathbb{R}_x)^{\times}$, where $\ket{f_n^{+}}\in \mathcal{S}_{-}^{\times}$ and $\ket{f_n^{-}}\in \mathcal{S}_{+}^{\times}$. Precisely, two Gelfand triplets are required to denote these spaces:
\begin{equation}\label{triplet}
    \mathcal{S}_{\pm}(\mathbb{R}_{x}) \subset \mathcal{L}^2(\mathbb{R}_{x}) \subset \mathcal{S}_{\pm}(\mathbb{R}_{x})^{\times},
\end{equation}
where $\mathcal{S_{+}} \cap \mathcal{S_{-}}=\left\{ \varnothing  \right\}$ and $\mathcal{S}=\mathcal{S}_{+}\cup\mathcal{S}_{-}$. 

The key focus of the IHO in this new basis, which will be crucial later, lies in the precise definition stated by Chruściński in Ref.~\cite{dariusz2} for the spaces $\mathcal{S}_{\pm}$. He demonstrates that when the solutions $X_{\pm}^{\mathscr{E}},\,Y_{\pm}^{\mathscr{E}}$ are extended to the complex plane, the complex eigenvalues $E_n$ appears as poles on the imaginary axis, while $f_n^{\pm}$ emerge as the residues. This outcome leads to the exact definition of the two Schwartz spaces in the following form:
\begin{equation}\label{space-definition}
\begin{aligned}
       \mathcal{S}_{-}&=\left\{\varphi \in \mathcal{S}(\mathbb{R}_{x}) \,\big| \bra{X_{\pm}^{\mathscr{E}}}\ket{\varphi}\in \mathcal{S}(\mathbb{R}_{\mathscr{E}})\cap\mathscr{H}^{2}_{-}(\mathbb{R}_{\mathscr{E}})\right\}\\
       \text{and} &\\
       \mathcal{S}_{+}&=\left\{\varphi \in \mathcal{S}(\mathbb{R}_{x}) \,\big| \bra{Y_{\pm}^{\mathscr{E}}}\ket{\varphi}\in \mathcal{S}(\mathbb{R}_{\mathscr{E}})\cap\mathscr{H}^{2}_{+}(\mathbb{R}_{\mathscr{E}})\right\},
\end{aligned}
\end{equation}
where $\mathscr{H}^2_{+}$ ($\mathscr{H}^2_{-}$) denotes the Hardy class space~\cite{Duren} for the upper (lower) half-plane. In other words, the space $\mathcal{S}_{+}$ represents well-behaved functions, which are boundary values of analytic functions in the upper half complex $\mathscr{E}$-plane, vanishing faster than any power of $\mathscr{E}$ at the upper semi-circle.  Meanwhile, $\mathcal{S}_{-}$ is the analogous for the lower complex $\mathscr{E}$-plane.

As a consequence, the two spaces $\mathcal{S}_{\pm}$ — which accommodate the two sets of solutions (denoted by $\pm$) — impose constraints on operators previously defined in the Hilbert space $\mathcal{H}$. Hence, this breaks the system's symmetry, as for each IHO, there exist two identical and indistinguishable states which are not symmetric. This constraint is notably manifested in the unitary time evolution operator $U(t)=e^{-i H_{\text{IHO}} t}$, which splits into two semigroups:
\begin{equation}
    \begin{aligned}
    &U_{+}(t)=U(t)\big|_{\mathcal{S}_{+}}: \mathcal{S}_{+} \to \mathcal{S}_{+} \quad \textrm{for}\quad t\leq 0,\\
    &\\
    &U_{-}(t)=U(t)\big|_{\mathcal{S}_{-}}: \mathcal{S}_{-} \to \mathcal{S}_{-} \quad \textrm{for}\quad t\geq 0.
    \end{aligned}
\end{equation}
As a result, the vectors $f_n^{+}\in \mathcal{S_{-}}(\mathbb{R}_{x})^{\times}$ exist only for $t\geq 0$, while $f_n^{-}\in \mathcal{S}_{+}(\mathbb{R}_{x})^{\times}$ exist only for $t\leq 0$, defining the so-called ``decay states'' $f_n^{+}(t,x)=U_{+}(t)f_n^{+}(x)$ and ``growth states'' $f_n^{-}(t,x)=U_{-}(t)f_n^{-}(x)$. The temporal reflection operator ($\mathrm{T}$) connects these states, such that $\mathrm{T}f_{n}^{\pm}(t,x)=f_n^{\mp}(t,x)$. Furthermore, in terms of decay theory, the mean life depends on the complex energy as $\tau\sim (n+1/2)^{-1}$, defining $f_0^{\pm}$ as the most ``stable'' state, i.e., the state with the largest mean life.

For completeness, let's highlight some properties of these generalized states $f_n^{\pm}$ that follow directly from the definitions (\ref{space-definition}):
\begin{enumerate}
    \item {\it They are conjugated to each other as}
    \begin{equation*}
        \overline{f_n^{\pm}(x)}=f_n^{\mp}(x);
    \end{equation*}
    \item {\it They are orthogonal}
    \begin{equation*}
        \bra{f_n^{\pm}}\ket{f_m^{\mp}}=\delta_{nm};
    \end{equation*}
    \item {\it They are complete in the following way}
    \begin{equation*}
        \mathbb{1}=\sum_{n=0}^{\infty}\ket{f_n^{\pm}}\bra{f_n^{\mp}}.
    \end{equation*}
\end{enumerate}

\subsection{Bound field as a collection of quantum IHOs}

Now, let's redirect our attention back to the bound state field $\phi_b(t,{\bf x})$ expressed in the form of Eq.~(\ref{phi_b1}) and quantize it by following a procedure similar to  the canonical quantization  of free fields. From the action (\ref{action-oscilator}), we have  a collection of inverted harmonic oscillators satisfying $|{\bf k}|<\gamma$.  Then we can anticipate that, at the quantum level, $\chi_{\bf k}(t)$ can be interpreted as an operator in the rigged Hilbert space.

By considering the transformation
\begin{equation}
    b_{\bf k}^{\pm}:=\sqrt{\frac{\omega_{\bf k}}{2}}\left(\chi_{\bf k}\pm\frac{1}{\omega_{\bf k}}\dot{\chi}_{\bf k}\right),
\end{equation}
with $\omega_{\bf k}=\sqrt{\gamma^2-{\bf k}^2}$, we quantize 
 the collection of IHO  by imposing the commutation relations
\begin{equation}\label{commutation2}
    \left[b_{\bf k}^{+},b_{{\bf k}^{\prime}}^{-}\right]=-i\delta({\bf k}-{\bf k}^{\prime})\quad\text{and}\quad \left[b_{\bf k}^{\pm},b_{{\bf k}^{\prime}}^{\pm}\right]=0.
\end{equation}
Each IHO mode solution satisfying $|{\bf k}|<\gamma$ (now characterized by $b^{\pm}_{{\bf k}}$) evolves in time with (\ref{hamiltonian3}) as its generator of time evolution. In the Heisenberg picture, this evolution can be found as
\begin{equation}\label{eq.heisenberg}
\frac{\mathrm{d}}{\mathrm{d}t}b^{\pm}(t)=-i \left[b^{\pm}(t),H_{\text{IHO}}\right] \Rightarrow b^{\pm}(t)=b^{\pm}e^{\pm \widetilde{\omega} t},
\end{equation}
where we have used the commutation relations (\ref{commutation}) and considered the initial condition $b^{\pm}(0)=b^{\pm}$. Similarly, for $x(t)$, we obtain
\begin{equation}\label{X_b}
x(t)=\frac{1}{\sqrt{2\widetilde{\omega}}}\left(b^{+}e^{\widetilde{\omega}t}+b^{-}e^{-\widetilde{\omega}t}\right).
\end{equation}

In a complete analogy, we expect that each IHO satisfying $|{\bf k}|<\gamma$ evolves in time like (\ref{X_b}), i.e., $\chi_{\bf k}(t)=\frac{1}{\sqrt{2\omega_{\bf k}}}\left(b_{\bf k}^{+}e^{\omega_{\bf k}t}+b_{\bf k}^{-}e^{-\omega_{\bf k}t}\right)$. Substituting this expression into (\ref{phi_b1}), the bound state field solution in the new base coordinates is finally given by
    \begin{equation}\label{phi_b_solution}
    \phi_b(t,{\bf x})=\sqrt{\gamma} \int \frac{\mathrm{d}^2{\bf k}}{2\pi}\frac{e^{i{\bf k}\underline{{\bf x}}-\gamma z}}{\sqrt{\omega_{\bf k}}}\left(b_{\bf k}^{+}e^{\omega_{\bf k}t}+b_{\bf k}^{-}e^{-\omega_{\bf k}t}\right).
\end{equation}
Here, we are just considering the time domain of $\phi_b(t,{\bf x})$ as a collection of inverted harmonic oscillators respecting their own Hamiltonian. The formal prescription will be given by the total Hamiltonian of the bound field as a sum over all individual IHOs (this will be discussed later in the next section). However, Eq.~(\ref{phi_b_solution}) shows an apparent classical time divergence that seems to persist in the $\phi_b(t,{\bf x})$ solution. Nevertheless, at the quantum level, the RHS always comes in pairs such that the action of the bound field on the quantum states will be constrained only for a specific domain of the time parameter $t$.

Therefore, let us introduce the {\it generalized ground states} $\ket{0^{\pm}}$, defined as
\begin{equation}
    b_{\bf k}^{\mp}\ket{0^{\pm}}=0,\quad \forall \, {\bf k}\quad \text{such that} \quad |{\bf k}|<\gamma.
\end{equation}
In direct analogy with the states $f_0^{\pm}$, these new states belong to the dual of the Schwartz space, i.e., $\ket{0^{\pm}}\in \mathcal{S_{\mp}^{\times}}$, defined by the triplet
\begin{equation}
    \mathcal{S}_{\pm}\subset \mathcal{H}\subset \mathcal{S}_{\pm}^{\times},
\end{equation}
where $\mathcal{S}_{\pm}$ are precisely defined by the relation (\ref{space-definition}). 

It is important to note that analogously to a complex field, the bound field $\phi_b(t,{\bf x})$ is associated with two kinds of states ($\pm$), with the subtle difference, the states $\ket{0^{+}}$ and $\ket{0^{-}}$ exist only for $t\geq 0$ and $t\leq 0$, respectively. In this way, we can define the one-quantum of decay (growth) state as $\ket{1_{\bf k}^{\pm}}=b_{\bf k}^{\pm}\ket{0^{\pm}}$. By applying $b_{\bf k}^{\pm}$ $n$-times, the state containing $n$-quanta with momentum ${\bf k}$ will be given by
\begin{equation}\label{states}
    \ket{n_{\bf k}^{\pm}}=\frac{1}{\sqrt{n_{\bf k}}}\left(b_{\bf k}^{\pm}\right)^n\ket{0^{\pm}},
\end{equation}
where $1/\sqrt{n_{\bf k}}$ is the normalization. As $t$ represents the time coordinate of the half-Minkowski space, the quantum solution $\phi_b(t,{\bf x})$ defines two kinds of ``particles'' in different time domains of the spacetime. As a consequence, the time reflection symmetry of $\overset{\circ}{\mathbb{H}}$ is not preserved by the bound solution — in agreement with the non-invariance of the solution (\ref{phi_b_solution}) when setting $t\to -t$. Since this solution originates from a non-positive self-adjoint extension of the operator $A$, it's not surprising that certain assumptions concerning dynamics, established in Ref.~\cite{waldishibashi}, are no longer applicable. Specifically, Assumption 2(ii) — time reflection invariance — does not hold for the bound state field solution.

In order to establish a Fock representation for the states (\ref{states}), we represent the $n$-fold tensor product of Hilbert spaces as $\mathcal{H}^{\otimes n}=\mathcal{H}\otimes \dots \otimes \mathcal{H}$, and then define the Fock space as the infinite direct sum
\begin{equation}
\mathscr{F}(\mathcal{H})\equiv\overset{\infty}{\underset{n=0}{\oplus}}\mathcal{H}^{\otimes n}.
\end{equation}
Similarly, for the space $\mathcal{S}$, we can represent the $n$-fold tensor product as $\mathcal{S}^{\otimes n}=\mathcal{S}\otimes \dots \otimes \mathcal{S}$, leading to the direct sum
\begin{equation}
    \mathscr{F}(\mathcal{S})\equiv \overset{\infty}{\underset{n=0}{\oplus}}\mathcal{S}^{\otimes n},
\end{equation}
which is understood as the union of $\mathscr{F}_{+}(\mathcal{S}_{+})$ and $\mathscr{F}_{-}(\mathcal{S}_{-})$. Considering the dual elements of these spaces, we can represent the {\it rigged Fock space} by the following triplet~\cite{gadella}
\begin{equation}\label{triplet2}
\mathscr{F}(\mathcal{S})\subset \mathscr{F}(\mathcal{H})\subset \mathscr{F}(\mathcal{S})^{\times}.
\end{equation}

Eq.~(\ref{groundstate}) ensures the existence of two {\it generalized Fock vacua}, which are elements of the larger Fock space $\mathscr{F}(\mathcal{S})^{\times}=\mathscr{F}_{+}(\mathcal{S}_{+})^{\times}\cup \mathscr{F}_{-}(\mathcal{S}_{-})^{\times}$. Thus, we can simple write $\ket{n_{\bf k}^{\pm}}\in \mathscr{F}(\mathcal{S})^{\times}$. Yet, it is possible now to define the {\it generalized ground state} for the total field $\Phi(t,{\bf x})=\phi(t,{\bf x})+\phi_b(t,{\bf x})$ as
\begin{equation}
    \ket{\Omega^{\pm}}=\ket{0}\otimes \ket{0^{\pm}}\in \mathscr{F}(\mathcal{S})^{\times},
\end{equation}
with $\ket{0}$ being the vacuum of the free field $\phi(t,{\bf x})$.

\section{The total conserved energy}\label{sec:stresstensor}

When dealing with fields in non-globally hyperbolic spaces such as $\overset{\circ}{\mathbb{H}}$ and  using the standard action formulation (i.e., the action containing only the bulk contribution), the total energy, derived from the stress-energy tensor, is not a conserved quantity over time. To address this issue, Wald introduced a new functional energy (see Eq. (11) in Ref.~\cite{waldishibashi}), which, in principle, is unrelated to the standard definition of the stress-energy tensor. However, in this section, we will recover the same conserved energy through the stress-energy tensor derived from the total action with the surface term (\ref{total-action}). This derivation remains valid even for fields with bound states.

\subsection{Stress tensor from the action and conserved energy}
To obtain the stress-tensor from the total action, we express the off-shell action explicitly in terms of its dependence on both the bulk metric $g_{\mu\nu}$ and the induced metric $h_{\mu\nu}$. This is achieved by rewriting Eq.~(\ref{total-action}) in the following manner
\begin{equation}
\begin{aligned}
    &S=-\frac{1}{2}\int_{\overset{\circ}{\mathbb{H}}}\mathrm{d}^4x\sqrt{|g|}g^{\mu\nu}\partial_{\mu}\Phi\partial_{\nu}\Phi+\frac{\gamma}{2}\int_{\partial\overset{\circ}{\mathbb{H}}} \mathrm{d}^3x\sqrt{|h|}\Phi^2.
\end{aligned}
\end{equation}

By taking the variation of the above action with respect to the metric tensor, we get
\begin{widetext}
    \begin{equation}\label{variation-of-S}
        \begin{aligned}
            \delta_gS&=-\frac{1}{2}\int_{\overset{\circ}{\mathbb{H}}}\mathrm{d}^4x\,\sqrt{|g|}\left(\partial_{\mu}\Phi\partial_{\nu}\Phi-\frac{1}{2}g_{\mu\nu}\partial^{\alpha}\Phi\partial_{\alpha}\Phi\right)\delta g^{\mu\nu}+\frac{\gamma}{2}\int_{\partial\overset{\circ}{\mathbb{H}}} \mathrm{d}^3x\,\left(-\frac{1}{2}\sqrt{|h|}h_{\mu\nu}\delta h^{\mu\nu}\right)\Phi^2\\
            &=-\frac{1}{2}\int_{\overset{\circ}{\mathbb{H}}}\mathrm{d}^4x\,\sqrt{|g|}\left(\partial_{\mu}\Phi\partial_{\nu}\Phi-\frac{1}{2}g_{\mu\nu}\partial^{\alpha}\Phi\partial_{\alpha}\Phi+\frac{1}{2}g_{\mu\nu}\delta(x;\partial \overset{\circ}{\mathbb{H}})\gamma\Phi^2\right)\delta g^{\mu\nu},
        \end{aligned}
    \end{equation}
\end{widetext}
where we first took the variation of the surface term with respect to $h_{\mu\nu}$ and then combined it with the bulk contribution by introducing the Dirac delta function $\delta(x;\partial\overset{\circ}{\mathbb{H}})$ defined as~\cite{saharian}
\begin{equation}
    \int_{\overset{\circ}{\mathbb{H}}}\mathrm{d}^{4}x\sqrt{|g|}\delta(x;\partial\overset{\circ}{\mathbb{H}})=\int_{\partial\overset{\circ}{\mathbb{H}}}\mathrm{d}^{3}x\sqrt{|h|}.
\end{equation} 

The classical stress tensor is conventionally defined in relation to the action $S$ as
\begin{equation}\label{stress-definition}
T_{\mu\nu}=\frac{-2}{\sqrt{|g|}}\frac{\delta S[\Phi,g_{\mu\nu}]}{\delta g^{\mu\nu}}.
\end{equation}
Comparing with Eq.~(\ref{variation-of-S}), we have
\begin{equation}\label{TEM}
        T_{\mu\nu}=\underbrace{\partial_{\mu}\Phi\partial_{\nu}\Phi-\frac{1}{2}g_{\mu\nu}\partial^{\alpha}\Phi\partial_{\alpha}\Phi}_{T_{\mu\nu}^{(V)}}+\underbrace{\frac{1}{2}g_{\mu\nu}\delta(x;\partial \overset{\circ}{\mathbb{H}})\gamma\Phi^2}_{T_{\mu\nu}^{(S)}}.
\end{equation}
In the above stress-tensor, two distinct contributions can be identified: one originating from the bulk action $T_{\mu\nu}^{(V)}$; and the other arising from the surface action $T_{\mu\nu}^{(S)}$, which is characterized by the Dirac delta function. The contribution $T_{\mu\nu}^{(V)}$ agrees with the standard definition, allowing us to understand the total stress-energy tensor $T_{\mu\nu}$ as an `improved' tensor, distinguishing it from the standard stress tensor by the surface term. Now, the energy density is written as
\begin{equation}
    T_{00}=\frac{1}{2}\left(\dot{\Phi}^2+\partial_i\Phi\partial^i\Phi\right)-\frac{1}{2}\delta(x,\partial\overset{\circ}{\mathbb{H}})\gamma \Phi^2,
\end{equation}
with $i=1,2,3$. That is, unlike the standard energy density, half-Minkowski space exhibits a flow density contribution from its timelike surface $\partial\overset{\circ}{\mathbb{H}}$. Thus, for the total energy we get
\begin{equation}\label{total-energy}
    E=\frac{1}{2}\int\mathrm{d}^3x\left[\dot{\Phi}^2+\partial_{i}\Phi\partial^{i}\Phi-\delta(x;\partial\overset{\circ}{\mathbb{H}})\gamma \Phi^2\right].
\end{equation}

The energy expression (\ref{total-energy}) plays a crucial role in comprehending the approach developed in this work. The surface action introduced in (\ref{total-action}) gives rise to an additional parabolic potential term $\frac{\gamma \Phi^2}{2}$ originating from the boundary of the space. Indeed, the boundary condition problem for the field at $z=0$ is analogous to the problem of a semi-infinite string with a boundary condition at the origin. In this scenario, the RBC characterizes a string coupled with a spring (with a negative constant $-\gamma$) at this point. Consequently, the total conserved energy of the system becomes a combination of ``string energy'' and ``spring energy.'' 

Similarly, for the total field, taking the derivative of $E$ with respect to time yields
\begin{equation}\label{de/dt}
\begin{aligned}
\frac{\mathrm{d}E}{\mathrm{d}t}&=\int\mathrm{d}^3x\left(\dot{\Phi}\ddot{\Phi}+\partial^i\Phi \dot{(\partial_i\Phi)}\right)-\int \mathrm{d}^2x\,\gamma \Phi \dot{\Phi}\\
&=\int\mathrm{d}^3x\,\dot{\Phi}\left(\ddot{\Phi}-\partial_i\partial^i\Phi\right)-\int \mathrm{d}^2x\,\dot{\Phi}\left(\Phi^{\prime}+\gamma \Phi\right),
\end{aligned}
\end{equation}
where we used integration by parts from the first to the second line. Utilizing Eq.~(\ref{EOM}), we immediately obtain $\mathrm{d}E/\mathrm{d}t=0$. Thus, the total energy is conserved over time.

It is crucial to note that in the absence of the surface action, the imposition of a boundary condition does not lead to conserved energy — except for the trivial Dirichlet boundary condition, which represents the particular case of $\gamma\to\infty$. Moreover,  Eq.~(\ref{total-energy}) provides a controlled flow of energy density at the boundary $\partial \overset{\circ}{\mathbb{H}}$,  induced by the surface action (\ref{total-action}).

\subsection{Expectation values}


We usually have to be careful when dealing with expected values of physical quantities that depend on quadratic forms of quantum fields — specifically, the quantity $\langle\psi|\Phi(t,{\bf x})^2|\psi\rangle$ becomes ill-defined at specific points in spacetime. As a consequence, even in the usual Minkowski spacetime, the Hamiltonian $H$ commonly exhibits ultraviolet divergence. To address this issue, various regularization methods can be applied to the expected value of the stress-tensor $\langle T_{\mu\nu}\rangle$ to obtain the regularized energy (e.g., cutoff function, zeta function regularization, point-splitting technique).

In this paper, the Hamiltonian can be decomposed into two components: $H=H^{(\text{free})}+H^{(\text{b})}$,  where $H^{(\text{free})}$ is the Hamiltonian associated with the free field and can be computed by directly inserting $\phi(t,{\bf x})$ into (\ref{total-energy}). On the other hand, $H^{(\text{b})}$ is the Hamiltonian for the bound field, obtained by inserting $\phi_b(t,{\bf x})$ into the same equation. In Ref.~\cite{casimir}, Romeo \& Saharian provided a detailed study of the regularization for the free field contribution. Precisely, as the divergence in this term comes from the bulk component of the stress tensor, and $\overset{\circ}{\mathbb{H}}$ is a flat space, the implemented regularization involves a direct subtraction of the Minkowski vacuum contribution from the volume term:
\begin{equation}
    \langle T^{(V)}_{\mu\nu}\rangle_{\text{reg}}=\langle T^{(V)}_{\mu\nu}\rangle-\bra{0_{M}}T^{(V)}_{\mu\nu}\ket{0_{M}},
\end{equation}
where $\ket{0_{M}}$ denotes the Minkowski vacuum. Subsequently, the obtained result can be combined with the surface term $\langle T_{00}^{(S)}\rangle$ to derive the corresponding energy. As this particular case was studied in detail in the aforementioned reference, we will concentrate on the contribution from the bound state.

By directly substituting (\ref{phi_b_solution}) into (\ref{total-energy}), we get the bound field Hamiltonian
\begin{equation}\label{H-bound}
    H^{(\text{b})}=-\int\mathrm{d}^2{\bf k}\,\omega_{\bf k}\left(b_{\bf k}^{\pm}b_{\bf k}^{\mp}\mp\frac{1}{2}\left[b_{\bf k}^{+},b_{\bf k}^{-}\right]\right),
\end{equation}
which characterizes an integration over all individual $H_{\text{IHO}}$ satisfying $|{\bf k}|<\gamma$.

Similar to the standard Hamiltonian in the Minkowski vacuum, the above expression has a divergence due to the commutation relation. This can be observed precisely by acting $H^{(\text{b})}$ on the ground states $\ket{0^{\pm}}$, yielding
\begin{equation}\label{groundstate-energy}
    H^{(\text{b})}\ket{0^{\pm}}=\left[\mp\frac{i}{2} \int\mathrm{d}^2{\bf k}\,\omega_{\bf k}\delta(0)\right]\ket{0^{\pm}},\quad |{\bf k}|<\gamma,
\end{equation}
which is characterized by the infinity $c$-number $\delta(0)$. If we consider placing the theory in a bi-dimensional box with sides of length $L$, we can interpret
\begin{equation}
    (2\pi)^2\delta(0)=\lim_{L\to\infty}\int_{-L/2}^{L/2}\int_{-L/2}^{L/2} \mathrm{d}^2x\,\left.e^{i {\bf k}\underline{{\bf x}}}\right|_{{\bf k}=0}=A,
\end{equation}
where $A$ is the box's area. Therefore, we recognize the remaining term in (\ref{groundstate-energy}) as the sum of ground state energies for each IHO, which is not divergent in the ultraviolet since $|{\bf k}|<\gamma$. However, it represents an infinite collection of zero-point energies for the IHO (spread over a surface with an infinite area).

In order to regularize the expectation value of $H^{(\text{b})}$ in any state, we  subtract the ground state divergent term from any state using the following expression (note the change from $\pm$ to $\mp$ to represent the transition from {\it ket} to {\it bra} elements)
\begin{equation}
    \langle H^{(\text{b})}\rangle_{\text{reg}}=\langle H^{(\text{b})}\rangle-\bra{0^{\mp}}H^{(\text{b})}\ket{0^{\pm}}.
\end{equation}
This procedure is equivalent to subtracting the commutator relation from (\ref{H-bound}). For the resulting Hamiltonian, we obtain $\left[H^{(\text{b})}_{\text{reg}},b_{\bf k}^{\pm}\right]=\mp i \omega _{\bf k}b_{\bf k}^{\pm}$. By substituting this into the Heisenberg equation, we generalize Eq.~(\ref{eq.heisenberg}). Furthermore, for an $n$-quanta of decay (growth) state, the regularized Hamiltonian yields the purely imaginary eigenvalue (using Eq.~(\ref{commutation2}))
\begin{equation}\label{imaginary-energy}
    H^{(\text{b})}_{\text{reg}}\ket{n_{{\bf k}}^{\pm}}=\mp i n_{\bf k} \omega_{\bf k}\Theta(\gamma-|{\bf k}|),\quad \pm t>0,
\end{equation}
where $\Theta$ is the Heaviside step function.

The above regularization procedure ensures that the eigenvalues are well-defined and avoids the divergence associated with the ground state terms. With this result, we can now fully interpret the meaning of the imaginary energy in a physical context. Let $\ket{\psi}$ denote any excited state of the free field $\phi(t,{\bf x})$ with energy $E_{0}$ (considering any regularization procedure already performed). Then, define the generalized state $\ket{\Psi^{\pm}} \in \mathscr{F}(\mathcal{S})^{\times}$ in the following manner
\begin{equation}
    \ket{\Psi^{\pm}}=\ket{\psi}\otimes\ket{g^{\pm}}.
\end{equation}
This ket element $\ket{\Psi^{\pm}}$ represents a state containing stable quanta with energy $E_0$ and $n$-quanta of decay (growth) states for $t>0$ ($t<0$). So, the action of the total regularized Hamiltonian in this new state is
\begin{equation}
    H_{\text{reg}}\ket{\Psi^{\pm}}=\left(E_0 \mp i \Gamma/2\right)\ket{\Psi^{\pm}},\quad \pm t>0.
\end{equation}

In the above equation, we have $\Gamma=2 n_{\bf k} \omega_{\bf k}\Theta(\gamma-|{\bf k}|)$. Essentially, the state $\ket{\Psi^{\pm}}$ represents the so-called Gamow vector~\cite{Gamow,Gentilini}. This vector is an eigenstate of the regularized Hamiltonian with a complex eigenvalue $E_0\mp i \Gamma/2$. Such states effectively encapsulate the resonance behavior of the free and bound field combined system, which in other scenarios is usually described by the well-established Breit-Wigner distribution (see Ref.~\cite{Resonance}). In simpler terms, each field state $\ket{\Psi^{\pm}}\in\mathscr{F}(\mathcal{S})^{\times}$ corresponds to a resonance (or an unstable state) around zero energy, where the free particle's energy $E_0$ characterizes the resonance energy. Meanwhile, the complex energy, which emerges from the bound field, defines the resonance width $\Gamma$, which, in turn, determines the system's mean life $\tau\sim1/\Gamma$, i.e., the characteristic time to the bound field change its state.

\section{Conclusion}\label{sec:conclusion}

In this paper, we explored the solutions of fields arising from non-positive self-adjoint extensions of the spatial part $A$ of the wave operator in the specific case of a scalar field propagating in (the non-globally hyperbolic) half-Minkowski space. The corresponding Robin boundary condition at $z=0$ gives rise to unstable classical solutions which were the focus of this work. Our main goal was the elucidation of the bound field as a set of mode states represented (in their time domain) as a collection of inverted harmonic oscillators — single particles being scattered by a parabolic barrier due to the potential originating in the timelike surface of $\overset{\circ}{\mathbb{H}}$.

Through the canonical quantization of this bound field, we demonstrated the quantum nature of the bound field, shedding light on its quantum aspects and interpreting it as an operator in the so-called rigged Fock space. As a result, we showed that the bound field is not invariant under time reversal, highlighting the non-trivial quantum behavior of this system, which defines two distinct ``particle'' states — growth and decay states. Specifically, this distinction is more apparent in the Schrödinger picture, where the bound field generates states at $t\to -\infty$ that grow in time until $t=0$. Subsequently, these states transform into ones that decay over time until they disappear at $t\to\infty$.

By investigating the energy for the prescribed dynamics (given the derivation of the energy functional from an action principle), we demonstrate that both the free and bound fields give rise to a conserved energy,  consistent with the Wald functional energy formulation. At the quantum level, we showed the regularized expected value of the bound Hamiltonian can be obtained by directly subtracting its expected value in the ground states, i.e., $\langle H^{(\text{b})}\rangle_{\text{reg}}=\langle H^{(\text{b})}\rangle-\bra{0^{\mp}}H^{(\text{b})}\ket{0^{\pm}}$. Consequently, the total regularized Hamiltonian, comprising contributions from both the free and bound fields, can be understood as the Hamiltonian of a resonance system, where its eigenstates $\ket{\Psi^{\pm}}$ represent Gamow vectors in the dual Fock space $\mathscr{F}(\mathcal{S})^{\times}$. Here, the regularized energy of the free field defines the resonance energy, while the complex energy of the bound field defines the state's mean life.

In future work, we aim to extend the results of this paper to any non-globally hyperbolic spacetimes. We will focus on examining the impact of reflection symmetry breaking in the bound state field on its causal propagator. Additionally, this exploration will enable us to understand the implications of resonant states through the response function of the Unruh-DeWitt detector model.

\acknowledgments

B. S. F. acknowledges support from the Conselho Nacional de Desenvolvimento Científico e Tecnológico (CNPq, Brazil), Grant No. 161493/2021-1. J. P. M. P. thanks the support provided in part by Conselho Nacional de Desenvolvimento Científico e Tecnológico (CNPq, Brazil), Grant No. 311443/2021-4, and Fundação de Amparo à Pesquisa do Estado de São Paulo (FAPESP) Grant No. 2022/07958-4. Additionally, we thank the anonymous referee for their invaluable comments and suggestions, which have greatly improved our manuscript.

\end{document}